\documentstyle[prl,twocolumn,aps,graphicx]{revtex}

\begin{document}

\draft

\title{On the scaling of the ($H$-$T$)-phase diagram of CuGeO$_3$.}

\author{J. Zeman$^{a}$\cite{byline}, G. Martinez$^{a}$,
P. H. M. van Loosdrecht$^{b}$, G. Dhalenne$^{c}$\ and A.
Revcolevschi$^{c}$}

\address{$^{a}$\ Grenoble High Magnetic Field Laboratory MPI-FKF/CNRS,
         25 Avenue des Martyrs, F-38042 Grenoble Cedex 9, France}
\address{$^{b}$\ II. Physikalisches Institut, RWTH-Aachen, Templergraben 55, D-52056
         Aachen,Germany}
\address{$^{c}$\ Laboratoire de Chimie des Solides, Universit\'{e} de Paris-Sud,
         b\^{a}timent 414, F-91405 Orsay, France}
\date{\today}
\maketitle

\begin{abstract}
The $H-T$\ phase diagram of CuGeO$_3$\ has been determined, for
different values of the hydrostatic pressure, utilizing optical
absorption spectroscopy on the Cu$^{2+}$\ $d-d$\ transitions. It
is shown that the intensity of the related zero phonon line
transition is very sensitive to the local environment of
Cu$^{2+}$, allowing for precise determination of all phase
transitions. It is found that the phase diagrams at various
pressures do not scale according to the Cross-Fischer theory. An
alternative scaling is proposed.
\end{abstract}
\vskip 5mm
Since the discovery of the spin-Peierls (SP) transition in the
inorganic compound CuGeO$_3$~\cite{HAS93} there has been an
increasing experimental and theoretical interest for the physics
related to low dimensional magnetism~\cite{BOU96}. One of the most
characteristic properties of a spin-Peierls material is its
magnetic field ($H$)-temperature ($T$) phase diagram (PD). For the
quasi-one-dimensional Cu$^{2+}$\ spin system CuGeO$_3$, this PD
has been explored using a variety of techniques
\cite{LOR98,REM,POI95}. The investigation using thermal
expansion/magnetostriction measurements \cite{LOR98,BUC96}
has shown that CuGeO$_3$, which
crystallizes into an orthorhombic structure ($a$=0.481 nm,
$b$=0.843~nm and $c$=0.295~nm), exhibits more drastic changes
along the $a$\ and $b$\ axis than along the $c$-axis (which bears
the Cu$^{2+}$\ ions) at all lines describing the PD. This
underlines the importance of the elastic energies involved in
these transitions.

From a theoretical point of view the corresponding PD is believed
to be universal~\cite{CRO79,CRO79A,BUZ98} defining in the $H-T$\
plane a high temperature uniform phase (U) and a dimerized (D)
phase below $T_{\rm sp}$\ and magnetic fields below a critical
value $H_{\rm c}$\ ($T_{\rm sp}\approx 14$~K and $H_{\rm c}\approx
13$~T for CuGeO$_3$). Above $H_{\rm c}$\ and at low temperatures
the system enters an incommensurate solitonic magnetic phase (I)
which evolves into a sinusoidally modulated phase for higher
fields~\cite{LOR98}. The best established theory for spin-Peierls
materials is the one due to Cross and Fisher
(CF)~\cite{CRO79,CRO79A}. Apart from the universality of the PD,
one of the main theoretical results is that, at low $H$\ values
the transition temperature scales with the square of the field.
For CuGeO$_3$, this scaling requires a further reduction of $H$\
by 10\% \cite{LOR97}. The universal character could be eventually
the consequence of approximations made in the model and since to
our knowledge there is no complete experimental study of this
aspect, the question deserves a careful inspection.

In order to achieve this goal one has to vary $T_{\rm  sp}$.
When accomplished by chemical
substitutions \cite{REN95,REG95,HAS95}, the question of
the relative quality and purity of samples becomes acute. A more
appropriate way is to apply, on a given sample, hydrostatic
pressure which increases $T_{\rm sp}$\ in CuGeO$_3$\
\cite{WIN95,LOO97}. It has been argued that this increase results
from an increasing frustration in the spin system under
pressure~\cite{BUC96,LOO97}. This, in turn, gives an additional
interest for the study of the PD under pressure. Technically
speaking the standard techniques to measure the PD are
not easy to settle in a high pressure cell. We have therefore
decided to adopt an optical technique which has been shown
recently to demonstrate around $\hbar \omega = 1.47$~eV
specific anomalies at the different phase transitions~\cite{LON97}.

After a brief description of the technical aspects, we will first
refine the analysis of the optical data trying to explain what is
indeed measured and how it can be interpreted. The second part of
this Letter presents and discusses the results obtained with this
technique under high pressure.

The crystals of CuGeO$_3$\ used in this study have been grown by a
traveling floating zone technique~\cite{REV93} and
cleaved in such a way the surface of the sample contains the $c$\
and $b$-axis. The absorption measurements in the visible region
around 1.47~eV have been done on samples with a thickness $d$=2~mm
using a standard optical set-up. Magnetic fields $H$\ up to
23~T were provided by a modified resistive Bitter magnet and
applied parallel to the $a$-axis of the sample. The high pressure
measurements were made in a small piston high pressure cell filled
with a mixture of ethanol-methanol, equipped with two sapphire
windows to perform transmission measurements. The temperature was
recorded with a cernox thermometer for which the
magneto-resistance has been calibrated at different temperatures.

The optical technique consists~\cite{LON97} in investigating the
optical properties of CuGeO$_3$\ in the low energy tail of the
intra-center absorption of the Cu$^{2+}$-ions. This absorption
band, centered at $\overline{E} \approx 1.63$~eV, has been
analyzed carefully \cite{BAS96,BES98}. It results in a broad band with a
maximum absorption coefficient $\alpha(\overline{E})$\ ranging
from 500 to 800 cm$^{-1}$\ (depending on the polarization), and
shows that in the range of 1.47~eV the remaining absorption
coefficient is expected to be less than 1 cm$^{-1}$. In these
conditions, the appropriate way to investigate optical
properties is not by measuring reflection, which becomes a
complicated function of reflectivity and transmission responses,
but by direct transmission measurements. If one performs such
experiments, on wedged samples to avoid interference effects, the
transmission $\mathcal{T}$\ is given by the standard expression
\begin{equation}
{\mathcal{T}}(\hbar \omega)=\frac{\left( 1-{\mathcal{R}} \right)^{2}
\exp\left(- \alpha(\hbar
\omega)d\right)}{1+{\mathcal{R}}^{2}\exp\left( -2\alpha(\hbar
\omega)d\right)}
\end{equation}
where ${\mathcal{R}}$\ is the reflectivity coefficient of one
interface. For low absorbing media ${\mathcal{R}}$\ reduces to
$(n-1)^{2}/(n+1)^{2}$, where $n$\ is the refractive index
($\sim 2.5$\ for CuGeO$_3$~\cite{BAS96}), implying ${\mathcal{R}}
\approx 0.13$. In such conditions, when measuring the ratio of the
transmission coefficient with respect to that of a reference point
(${\mathcal{T}}_0$), the denominator of Eq.~1 can be approximated
to 1. Furthermore, since ${\mathcal{R}}$\ is not expected to vary
significantly neither with $T$\ nor $H$\ one arrives at
${\mathcal{T}}/ {\mathcal{T}}_0= \exp\left( -(\alpha - \alpha_{0})
d \right) = \exp(- \Delta \alpha \ d)$, where all optical
functions are dependent on the photon energy $\hbar\omega$. The
absorption strength is strongly polarized~\cite{LON97,BAS96}, but
all discontinuities related to the phase transitions are not
dependent on this polarization~\cite{LON97}. Since it is not easy
to perform polarized measurements in a high pressure cell, we
report here on non-polarized measurements only. The relevant
physical quantity is indeed $\Delta\alpha d = -\ln
({\mathcal{T}}/{\mathcal{T}}_{0})$. A typical variation of this
quantity at fixed temperature and different values of $H$\ is
presented in Fig.~1b. One sees a clear discontinuity of the
intensity, corresponding to the first-order transition at the
critical field $H_{\rm c}$\ as already reported~\cite{LON97}. We
shall see below that this value corresponds quite well to $H_{\rm
c}$\ values reported by other techniques.

Before detailing the main results one has to discuss the origin of
the optical transition we are studying. In the picture of
non-interacting ions the absorption occurs between vibronic
transitions involving as the initial state an electron in the $d$\
fundamental state and $m=0$\ vibrational state (at low $T$) and as
the final state the electron in the $d^{*}$\ excited state of the
ion and $m=l$\ phonons in a displaced configuration coordinate
$Q_r$\ with respect to the initial value $Q_0$. The system relaxes
in the excited state, after absorption, towards a lower energy by
emitting these $l$\ phonons~\cite{STU67}. The strength of the
optical matrix element is essentially governed by the overlap of
the $m=0$\ phonon wave function  and the displaced $l$\ phonon
wave function. For simple cases where only one type of phonons is
involved one gets the picture sketched in Fig.~1a. Therefore
one observes around 1.47~eV the so-called
zero-phonon-line (ZPL) for which $l=0$\ and the energy is
$E_{ZPL}$. Of course due to the coupling of the different
Cu$^{2+}$\ ions, one expects the corresponding transition be
broadened and this is indeed observed since the width $\Gamma$\ of
the spectra is of the order of 10~meV and does not vary much with
$T$\ and $H$. Because this width remains much smaller than that of
the main absorption band ($\approx0.5$~eV)~\cite{BAS96} one may
still apply, in a first approximation to our case, the results
known for isolated centers. In this context the probability to
absorb a photon in the $l$th vibronic state, neglecting the
constant electronic part of the matrix element, is given by:
$P(l)=\exp (-S) S^{l}/l!$. $S$\ is the number of phonons emitted
after the absorption at $\overline{E}$ (Fig.~1a).
The phonon involved in this
process has been determined in Ref.~18 and corresponds
very likely to the IR active phonon at $\hbar\omega_{p}=
28.5$~meV~\cite{POP95A}. With that analysis one gets $S$ values
of about 6 which correspond to a very strong electron-phonon
interaction. This gives in turn a mean absorption coefficient of
the ZPL of about 0.25~cm$^{-1}$\ which justifies the transmission
technique used here. In non-polarized experiments the situation is
not simple because, if there exists a nice well-defined line for
the electric field of the light $\vec{E}
\parallel c$-axis, the situation for $\vec{E}\parallel b$-axis is
more complex~\cite{LON97} and this explains the high energy
shoulder of the spectra in Fig.~1b. Therefore $S$\ has to be
understood as an averaged value of the relaxation energy of the
Cu$^{2+}$ions in the excited state over the different relaxation
routes. The total integrated quantity, $M_{0}(H,T)=d \int
\Delta\alpha \ dE$, has been evaluated over a fixed range of
energies between 1.45~eV and 1.50~eV. If one knows $M_0$\ at a
reference point, which will be taken for each pressure at $T=2$~K
and $H=0$~T, it is easy to show that the variation of $M_0$\
mimics quite well that of $\Delta S =S(0,2~K)- S(H,T)$\ {\it i.e.}
the variation of the relaxation energy with $H$\ and/or $T$.
\begin{figure}[h]
\begin{center}
\includegraphics[width=6cm,clip=true]{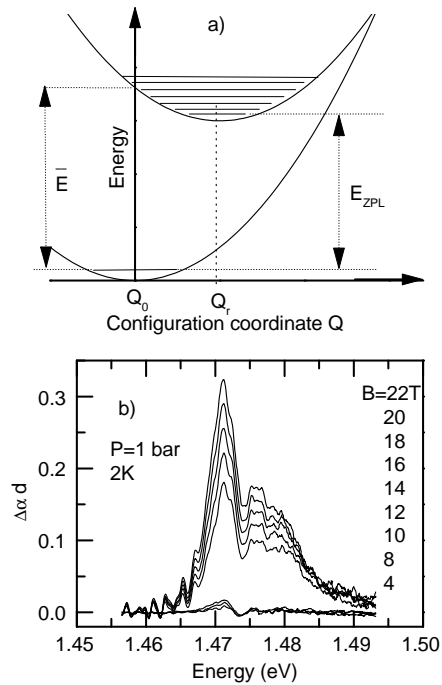}
\end{center}
\label{fig1} \caption{a) schematic configurational diagram for the
intra-center absorption of the Cu$^{2+}$ions; b) variation of
$\Delta \alpha d$\ at $P=1$~bar, $T=2$~K for different values of
magnetic field increasing by steps of 2~T beyond 8~T.}
\end{figure}
\begin{figure}[h]
\begin{center}
\includegraphics[width=6cm,clip=true]{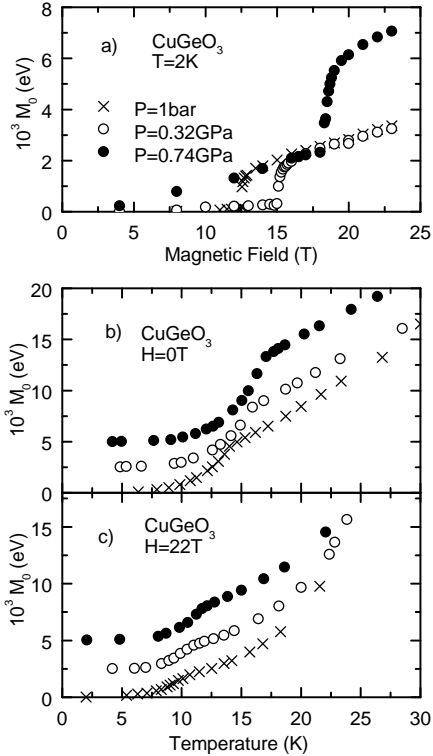}
\end{center}
\label{fig2} \caption{ a) Variation of $M_{0}$\ at $T=2$~K as a
function of $H$\ for different values of the pressure. b) Variation of
$M_{0}$\ at $H=0$~T as a function of $T$\ for different values of
$P$. c) Variation of $M_{0}$\ at $H=22$~T as a function of
$T$\ for different values of $P$. The $M_{0}$\ values  have
been shifted by 0.0025 for $P=0.32$~GPa and by 0.005 for
$P=0.74$~GPa in b) and c).}
\end{figure}
Fig.~2 presents characteristic variations of $M_{0} (H,T)$\ for
pressures $P=1$~bar, 0.32~GPa, and 0.74~GPa. The field dependence
at fixed temperature ($T=2$~K , Fig.~2a) shows an abrupt
discontinuity of $M_0$\ at a critical magnetic field $H_{\rm
c}(P)$. This discontinuity jump decreases upon increasing $T$\ and
disappears beyond some temperature $T^{*}(P) < T_{\rm sp}(P,H=0)$.
The existence of this discontinuity is related to the first-order
phase transition from the dimerized (D) phase to the
incommensurate (I) phase. Note that the discontinuity also
decreases at fixed temperature when increasing $P$. The
temperature dependence of $M_0$\ at fixed magnetic field $H=0$~T
(Fig.~2b) and 22~T (Fig.~2c) corresponding to fields lower and
higher than $H_{\rm c}$, respectively for all values of $P$) shows
a clear discontinuity in the slope $dM_0/dT$\ at a critical value
$T_{\rm c} (H, P)$. This discontinuity corresponds to the
second-order phase transition from the D to the U phases for $H<H_{\rm
c}$\ and from the I to the U phases for $H>H_{\rm c}$. It is
interesting to compare all the variations of $M_0$\ observed with
those of the lattice parameters of the structure. It has been
shown for instance~\cite{LOR98,BUC96} that, at $H=0$~T, $db/b$\
increases by about $8 \times 10^{-5}$\ between 2~K and 14~K
whereas $da/a$\ decreases by $5\times 10^{-5}$. These variations
constrain the environment of the Cu$^{2+}$\ ions which in turn is
expected to reduce the relaxation energy of these ions in the
excited state. This is indeed observed since in that range of
temperature $S$\ is found to increase by a factor of about 2
corresponding to a decrease of the relaxation energy of about 57~meV.

Now that the nature of the 1.47~eV absorption feature in
CuGeO$_3$\ has been established, we turn to the main results
reported in this Letter. Using a CF type of scaling ({\it i.e.}
$T/T_{\rm sp}$\ vs. $g\mu_bH/T_{\rm sp}$\ where $g=2.15$\ for
$H\parallel a$-axis \cite{PIL97}), the scaled PD of CuGeO$_3$\ is
shown in Fig.~3a for various pressures together with already
published data (small squares) \cite{LOR98}. Clearly one
reproduces these earlier results quite well at $P=1$~bar. The
transition temperatures and fields used in the scaling are listed
in Table~\ref{table1}. At low magnetic fields, the D--U transition
temperature indeed scales with the square of the
field~\cite{CRO79}. At higher fields, and in particular for the
D-I transition, the scaling no longer holds. Already present at
$P=1$~bar, the problem becomes more acute at higher pressures
(Fig.~3a). In order to reconcile the theoretical result with the
data, one now has to assume a $T_{\rm sp}$-dependent field
scaling factor, which is not very useful if one tries to unify the
PD. It seems therefore more appropriate to look for a different
kind of scaling law. It is here proposed that $H$\ does not scale
with $T_{\rm sp}$, but rather with the singlet-triplet gap
$\Delta_{\rm ST}$ at zero field. The pressure dependence of
$\Delta_{\rm ST}$\
has been determined by Nishi {\it et al.}~\cite{NIS95A}:
$\Delta_{\rm ST}(P)=\Delta_{\rm ST}(0)+10.5P$\ ($P$ in GPa,
$\Delta_{\rm ST}$\ in cm$^{-1}$, $\Delta_{\rm
ST}(0)=16.8$~cm$^{-1}$). The resulting scaled PD is shown in
Fig.~3b. Clearly the PDs at various pressures now collapse onto
each other, despite the observation that the D--I transition
occurs before the full closure of the magnetic gap.

In addition to the scaling properties of D--I transition, the
scaling of the D-U transition has also improved. The quadratic
field dependence found for the D-U transition is:
\begin{equation}
\frac{T_{\rm sp}(H)}{T_{\rm sp}(0)}
=\left(1-0.384\cdot\left[\frac{g\mu_bH}{\Delta_{\rm
ST}}\right]^2\right)
\end{equation}
This scaling is not compatible with the CF type of scaling and
appears more compatible with the mean field BCS type theory. However,
CuGeO$_3$\ does not obey this theory since
$\Delta_{ST}/T_{sp}$\ is not constant as can be inferred from
Table~1. As a consequence, one may conclude that the CF theory is
not very appropriate to describe the properties of CuGeO$_3$.
There are various factors which could explain that, in particular,
the presence of a next-nearest neighbor frustrating interaction
\cite{RIE95} or the non-adiabacity \cite{GOT98} (the exchange
energy is comparable to the phonon energies). It is interesting to
see whether the present scaling also applies to organic
spin-Peierls compounds such as (TTF)Cu(BDT)\ \cite{BLO81},
(TTF)Au(BDT) \cite{No82} and MEM-(TCNQ)$_2$\ \cite{BLO80} for
which, to our knowledge, $\Delta_{ST}$\ has not been directly
measured. However, if we speculate for them a value of
$\Delta_{ST}/T_{sp} \approx 1.76$\ (mean field theory), the
proposed scaling is surprisingly good for the D--U and even the
D--I phase transitions. In particular $g\mu_{B}H_{c}/\Delta_{ST}$\
ranges between 0.77 and 0.79 for all organic compounds. The
surprise comes from the fact that all compounds listed in Table~1
have very distinct properties concerning the existence of
frustration, soft phonon modes or variation of $T_{sp}$ with
pressure. This result implies that $H_c$ does not scale with
$\Delta_{ST}$ as already pointed by Ohta et al. \cite{Ohta}.
\begin{figure}[h]
\begin{center}
\includegraphics[width=6cm,clip=true]{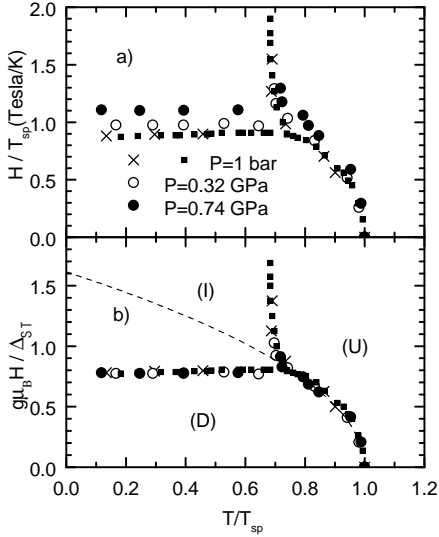}
\end{center}
\label{fig3} \caption{Phase diagram of CuGeO$_3$\ for different
pressures plotted versus $T/T_{sp}$; a) scaling of $H$\ with
$T_{sp}$\ and b) scaling of $H$\ with $\Delta_{ST}$. The solid
squares are data obtained at 1~bar by magnetostriction/thermal
expansion measurements$^3$. The dash line represents the variation
given by Eq.~2.}
\end{figure}

In conclusion the study of the ZPL absorption of the intra-center
Cu$^{2+}$\ transition has allowed to determine the $H-T$\ phase
diagram of CuGeO$_3$\ for different pressures. It has been shown
that a Cross-Fischer type of scaling of these phase diagrams is
not appropriate. It is noted that a scaling which does not assume
a constant ratio between $\Delta_{ST}$\ and $T_{sp}$\ could be
more effective, although it relies on two independent parameters
$\Delta_{ST}$\ and $T_{sp}$. We have at present no strong
arguments which could support these findings but we hope that this
observation will motivate further theoretical research on this
special class of compounds.

\acknowledgments

The Grenoble High Magnetic Field Laboratory is ``Laboratoire
conventionn\'{e} \`{a} l'UJF et l'INPG de Grenoble''. The Laboratoire
de Chimie des Solides is a "Unit\'{e} de Recherche Associ\'{e}e
CNRS n$^{\circ}$ 446".
J. Z. acknowledges the partial supports from the
grant ERBCHGECT 930034 of the European Commission.

\begin{table}
\caption{\label{table1} Measured phase diagram parameters}
\begin{tabular}{lllll}
Compound &P (GPa) &$T_{\rm sp}$~(K)&$\Delta_{ST}$~(cm$^{-1}$)&$H_c$~(T)\\
\hline
 CuGeO$_3$           &   0  &  14.2 &  16.8 &   12.9\\
                     & 0.32 &  15.5 &  20.3 &   15.1\\
                     & 0.74 &  17.0 &  24.7 &   18.8 \\
\hline MEM-(TNCQ)$_2$ & 0    &  18   &   ?   &   19.2 \\
\hline (TTF)CuS$_4$(CF$_3$)$_4$& 0     &  11 &   ?   & 11.6\\
\hline (TTF)AuS$_4$(CF$_3$)$_4$& 0     &  2.0  &   ?   & 2.1
\end{tabular}
\end{table}

\begin{references}
\bibitem[*]{byline}
On leave from the Institute of Physics, Academy of Sciences of Czech Republic,
  Prague.
\bibitem{HAS93}
M. Hase, I. Terasaki, and K. Uchinokura, Phys. Rev. Lett. {\bf 70},  3651
  (1993).
\bibitem{BOU96}
J.~P. Boucher and L.~P. Regnault, J. Phys. I {\bf 6},  1939  (1996).
\bibitem{LOR98}
T. Lorenz {\it et~al.}, Phys. Rev. Lett. {\bf 81},  7749  (1998).
\bibitem{REM}
G. Rem\'enyi {\it et~al.}, J. Low Temp. Physics {\bf 107}, 243 (1997).
\bibitem{POI95}
M. Poirier {\it et~al.}, Phys. Rev. B {\bf 51}, 6148, (1995).
\bibitem{BUC96}
B. B{\"u}chner {\it et~al.}, Phys. Rev. Lett. {\bf 77},  1624  (1996).
\bibitem{CRO79}
M.~C. Cross and D.~S. Fisher, Phys. Rev. B {\bf 19},  402  (1979).
\bibitem{CRO79A}
M.~C. Cross, Phys. Rev. B {\bf 20},  4606  (1979).
\bibitem{BUZ98}
A. Buzdin, H. Kachkachi, and Y. Meurdesoif, Physics Lett. A {\bf 237},  276
  (1998).
\bibitem{LOR97}
T. Lorenz {\it et~al.}, Phys. Rev. B {\bf 55},  5914  (1997).
\bibitem{REN95}
J.~P. Renard {\it et~al.}, Europh. Lett. {\bf 30},  475  (1995).
\bibitem{REG95}
L.~P. Regnault {\it et~al.}, Europhys. Lett. {\bf 32},  579  (1995).
\bibitem{HAS95}
M. Hase {\it et~al.}, Physica B {\bf 215},  164  (1995).
\bibitem{WIN95}
H. Winkelman {\it et~al.}, Phys. Rev. B {\bf 51},  12884  (1995).
\bibitem{LOO97}
P.~H.~M. van Loosdrecht {\it et~al.}, Phys. Rev. Lett. {\bf 79},  487  (1997).
\bibitem{LON97}
V. Long {\it et~al.}, Phys. Rev. B {\bf 56},  R14263  (1997).
\bibitem{REV93}
A. Revcolevschi and G. Dhalenne, Adv. Mater. {\bf 5},  9657  (1993).
\bibitem{BAS96}
M. Bassi {\it et~al.}, Phys. Rev. B {\bf 54},  R11030  (1996).
\bibitem{BES98}
B.~Beschoten, P.H.M. van Loosdrecht, unpublished (1998).
\bibitem{STU67}
M.~D. Sturge,  in {\em Solid state physics}, edited by F. Seitz and D. Turnbull
  (Academic Press, New York, 1967), Vol.~20, p.\ 178.
\bibitem{POP95A}
Z. Popovi{\'{c}} {\it et~al.}, Phys. Rev. B {\bf 52},  4185
(1995).
\bibitem{PIL97}
B. Pilawa {\it et~al.}, J. Phys.: Condens. Matter {\bf 9},  3779  (1997).
\bibitem{NIS95A}
M. Nishi {\it et~al.}, Phys. Rev. B {\bf 52},  R6959  (1995).
\bibitem{RIE95} J. Riera and A. Dobry, Phys. Rev. B {\bf 51},
16098 (1995); G. Castilla {\it et~al.}, Phys. Rev. Lett. {\bf 75},
1823 (1995).
\bibitem{GOT98}
G. S.~Uhrig, Phys. Rev. B {\bf 57}, R17004 (1998).
\bibitem{BLO81}
D. Bloch {\it et~al.}, Phys. Lett. A {\bf 82},  21  (1981).
\bibitem{No82} J. A. Northby {\it et~al.}, Phys. Rev. B {\bf 25},
3215, (1982).
\bibitem{BLO80}
D. Bloch {\it et~al.}, Phys. Rev. Lett. {\bf 44},  294  (1980).
\bibitem{Ohta} H. Ohta {\it et~al.}, J. Phys. Soc. Jpn. {\bf 63},
2870, (1994).
\end{references}
\end{document}